\documentclass[11pt]{iopart}
\usepackage{graphicx,epsf}
\usepackage{bm}

\begin{document}

\vspace{2cm}
\title{Splitting of critical energies in the $n$=0 Landau level of graphene}
\vspace{2cm}
\author{Ana L C Pereira}

\address{Instituto de F\'{\i}sica, Universidade Estadual de Campinas -
C.P. 6165, 13083-970, Brazil}
\ead{analuiza@ifi.unicamp.br }

\begin{abstract}

The lifting of the degeneracy of the states from the graphene $n$=0 Landau level (LL) is investigated through a non-interacting tight-binding model with random hoppings.  A disorder-driven splitting of two bands and of two critical energies is observed by means of density of states and participation ratio calculations. The analysis of the probability densities of the states within the $n$=0 LL provides some insights into the interplay of lattice and disorder effects on the splitting process. An uneven spatial distribution of the wave function amplitudes between the two graphene sublattices is found for the states in between the two split peaks.
It is shown that as the splitting is increased (linear increasing with disorder and square root increasing with magnetic field), the two split levels also get increasingly broadened, in such a way that the proportion of the overlapped states keeps approximately constant for a wide range of disorder or magnetic field variation.

\end{abstract}


\maketitle

\section{Introduction}

Since the first observation of the anomalous integer quantum Hall effect in graphene \cite{novoselov,zhang1}, there has been an enormous interest in the quantum Hall regime of graphene.
The sequence of quantized Hall plateaus in graphene is shifted by half-integer if compared to the well-known quantum Hall effect in usual two-dimensional electron gases. This is understood to be due to the unique features of the Dirac-like band structure of graphene around Fermi energy \cite{reviews}, which give origin to the following Landau level (LL) quantization: $E_n=\pm \nu_{F} \sqrt{2e\hbar B |n|}$, where $n$ is the Landau level index and $\nu_F$ is the Fermi velocity. In this way, the energy dependence for the LLs with magnetic field $B$ is not linear anymore, but goes with the square root of $B$ (positive energies for electrons and negative energies for holes), and there is also a LL at zero-energy (the $n$=0 LL), which is shared by electrons and holes.
In addition to the spin degree of freedom, graphene LLs exhibit a valley (sublattice) degree of freedom, giving a fourfold degeneracy to each LL.


Of special interest in the recent literature is the question of the lifting of the $n=0$ LL degeneracies, observed experimentally \cite{zhang2,jiang,ong,andrei,giesbers} and discussed in many theoretical works [9-22]. Amongst these works, there are various different proposed explanations for the origins of the observed splittings, however a consensus is still missing. Two of the experimental works \cite{zhang2,jiang} show a complete lifting of the $n$=0 fourfold degeneracy occurring for high magnetic fields up to 45T and low temperatures, and there are evidences that the extra quantum Hall plateaus observed at filing factors $\nu=0$ and $\nu=\pm 1$ should be due to the lifting of the spin and sublattice degeneracy, respectively  \cite{jiang}. The other three experiments \cite{ong,andrei,giesbers} show the lifting of only one of the degeneracies. The authors from ref. \cite{giesbers} suggest that the opening of the gap in the $n$=0 observed by them might be spin related (Zeeman splitting), because of a linear variation of the gap with $B$, however, in ref. \cite{andrei} the authors believe that the gap they measure should be driven by a broken sublattice symmetry. The physical mechanisms leading to the sublattice symmetry breaking are still not completely understood.

In this work, a splitting of two critical energies in the $n$=0 LL is observed within a nearest-neighbor tight-binding model for the graphene lattice, considering random hoppings as the only source of disorder. The splitting is inferred by means of participation ratio calculations, which show clearly the presence of two split peaks indicating two critical energies inside the $n$=0 Landau band, with localized states in between the peaks, around $E$=0. The splitting is found to have a linear dependence with the disorder strength and a square root dependence with magnetic field, in agreement with the numerical calculations from ref. \cite{schweitzer} and with the magnetic field dependence observed for the gap in ref. \cite{jiang}. As the model used in this work considers a single particle picture (no electron-electron interaction is taken into account) and there are no spin-dependent terms in the Hamiltonian, the splitting observed is related only to the effects of the random hopping disorder, which promotes intervalley mixing. To help in the understanding of these effects, I look here to the wave functions (probability densities) of the states within the $n$=0 LL, inspecting in particular the distribution of amplitudes in each of the sublattices, which gives interesting information about the process of disorder-driven sublattice mixing.

A unit cell of the graphene hexagonal lattice contains two carbon atoms, defining two sublattices, $A$ and $B$. For the perfect graphene lattice (in absence of any kind of disorder), the wave functions of all the states within the $n$=0 LL have non-zero amplitudes in only one of the sublattices. ``How does disorder affect this picture?" was the question addressed in Ref.\cite{ana_valley}, where it is shown how the increasing inclusion of a diagonal disorder (both white-noise or a Gaussian-correlated disorder model)   increasingly promotes the spread of the wave functions to both sublattices in the $n$=0 LL, a effect that revealed to be directly connected to an anomalous localization: an enhancement of the participation ratio of the states with increasing disorder \cite{ana_valley,ana_hmf18}. However, for diagonal disorder models (on-site energy fluctuations) the splitting of critical energies does not occur. Here the same question is addressed for a off-diagonal disorder model, for which a completely different picture emerges. The probability densities of the states from the $n$=0 LL are not anymore concentrated predominantly in one of the the sublattices, even when small disorders are considered. However, the  interesting find is that for the states closest to $E$=0 (which get localized for the present disorder model, contributing to the critical energies splitting process), the spatial distribution of the wave functions in one sublattice is completely different and independent of the distribution in the other sublattice. These are states corresponding to the region where the two split bands have an important overlapping.

\section{Model: Tight-Binding Hamiltonian with Random Hopping}

The tight-binding Hamiltonian model for the honeycomb lattice used in this work is defined by:

\begin{equation}
H =  \sum_{<i,j>} (t_{ij}e^{i\phi_{ij}} c_{i}^{\dagger} c_{j} + t_{ij}e^{-i\phi_{ij}}
c_{j}^{\dagger} c_{i})
\end{equation}

\hspace{-\parindent}where $c_{i}$ is the fermionic operator on site $i$ and the sum runs over nearest-neighbor bonds. The hopping parameters $t_{ij}$ randomly fluctuate (white-noise) from bond to bond around the average value $t$:

\begin{equation}
t - \frac{W}{2} \leq t_{ij} \leq t + \frac{W}{2}
\end{equation}

Thus $W$ represents the width of the uniform off-diagonal disorder distribution. $t$$\approx$2.7eV for graphene, however note that the results shown here are parameterized by $t$.
The graphene lattices  considered have $M$$\times$$N$ atoms ($M$ zig-zag chains, each containing $N$ atoms), which are repeated by means of periodic boundary conditions. In this way, the dimensions of the lattices that are periodically repeated are given by:

\begin{equation}
 L_x=(M-1)a\frac{\sqrt{3}}{2}, \;\;\;\;\,\,\,\,\, L_y=(N-1)\frac{a}{2},
\end{equation}

\hspace{-\parindent}where $a$=2.46{\AA} is the lattice constant for graphene.

The perpendicular magnetic field $B$ is included by means of a Peierls' substitution (a complex phase factor in the hopping parameter): $\phi_{ij}= 2\pi(e/h) \int_{j}^{i} \mathbf{A} \! \cdot \! d \mathbf{l} \;$. In the Landau gauge, $\phi_{ij}\!=\!0$ along the $x$ direction and $\phi_{ij}\!=\pm \pi (x/a) \Phi / \Phi_{0}$ along the $\mp y$ direction, with $\Phi / \Phi_{0}=Ba^{2}\sqrt{3}e/(2h)$. The focus is on the low-flux limit ($\Phi / \Phi_{0} < 0.05$), where the graphene LLs are well defined in this model.

\section{Density of States and the n=0 LL Splitting}

\begin{figure}[b]
\vspace{0.3cm}
\includegraphics[width=15.4cm]{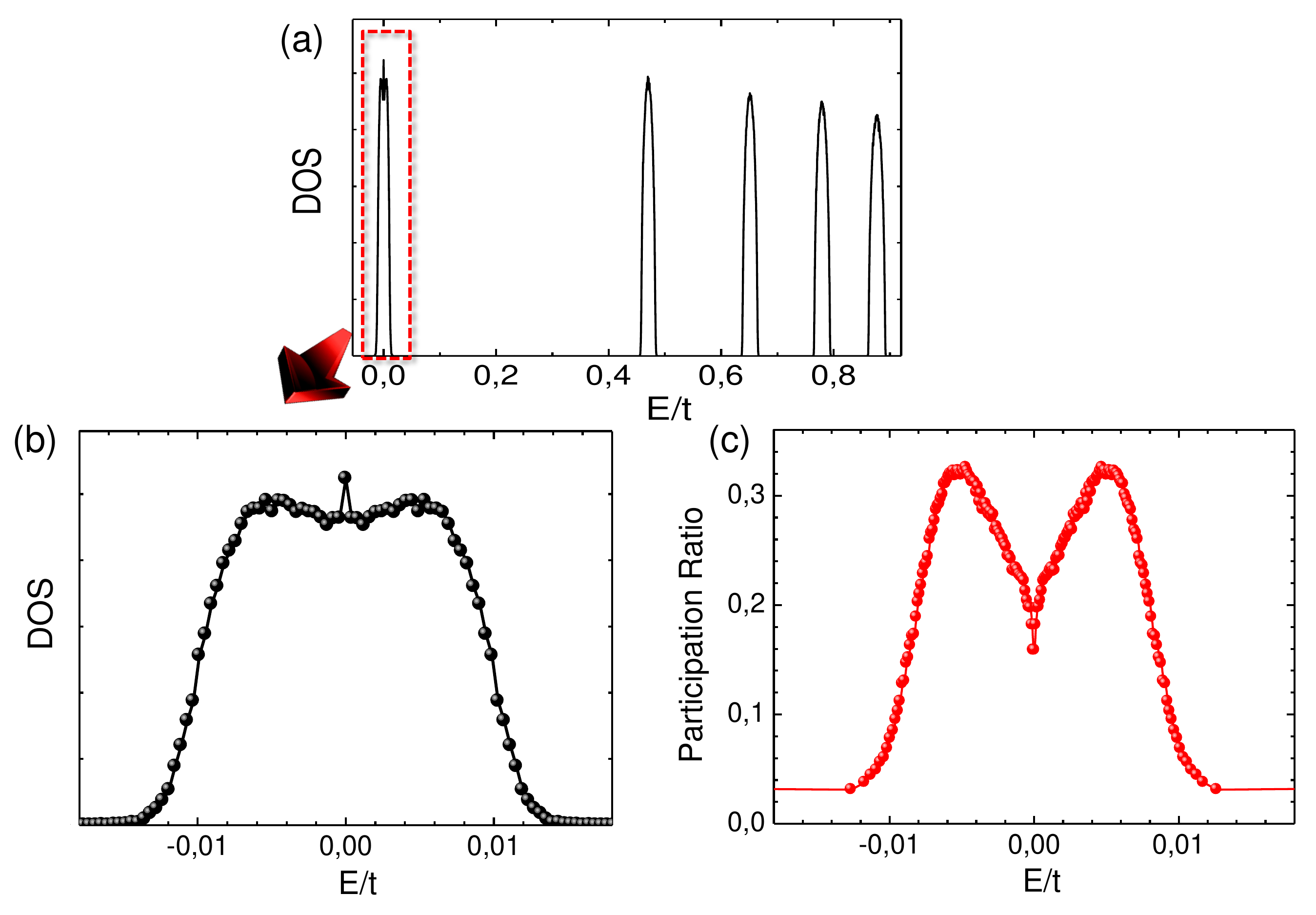} \hspace{-0.5cm}
\caption{{\bf (a)} Density of states showing five Landau levels (from $n$=0 to $n$=+4), broadened by a random hopping disorder with $W/t=0.1$, averaged over 360 disorder realizations. The size of the lattice considered is 19.8nm$\times$11.4nm ($M$$\times$$N$=94$\times$94), with magnetic flux $\Phi/\Phi_0=2/94\approx 0.021$. {\bf (b)}  Zoom showing details of the shape of the DOS for the $n$=0 LL, including the small peak at E=0. {\bf (c)} The corresponding participation ratio of the states from the $n$=0 LL, calculated for the same parameters, also averaged over 360 disorder realizations. A clear splitting in two critical energies can be observed.}
\end{figure}

In Figure 1(a) one can see the density of states (DOS) with the formation of five Landau levels, $n$=0 to $n$=+4, corresponding to a magnetic flux $\Phi/\Phi_0=2/94\approx 0.021$. The broadening of the levels are due to a random hopping disorder strength of $W/t=0.1$. The lattice considered has $M$$\times$$N$=94$\times$94, which means dimensions of $L_x$=19.8nm along the armchair boundary and $L_y$=11.4nm along the zigzag boundary, defining the disordered unit cell that is periodically repeated (periodic boundary conditions). The spectrum is symmetric about zero energy, thus the $n<0$ Landau levels are not shown here. The DOS is an energy histogram accumulated after the Hamiltonian diagonalization for hundreds of disorder realizations - in the case of Fig.1, 360 realizations were taken.
Here we see that the $n$=0 LL has about the same broadening width of the higher levels, which is different from the result obtained in Ref.\cite{schweitzer} for a random magnetic flux disorder. However, the dependence of the $n$=0 LL broadening with the correlation length of the random hopping  was already elucidated in Ref.\cite{aoki}, which shows a DOS with broadenings in accordance to the results shown here for spatially uncorrelated random bonds.

Fig.1(b) shows a zoom of the DOS for the region of the $n$=0 LL, allowing the observation in more details of the shape of this central band, which has many differences when compared to the $n$=0 Landau band shape obtained for a diagonal disorder model\cite{ana_valley,ana_hmf18} and even when compared to the higher levels. One can see in Fig1(b) that the broadening of this central LL does not resemble the usual Gaussian or semi-elliptic LL line shape, but instead, has a lowering in the density of states around the center (apart from the presence of a small peak observed in the DOS for the states closer to $E=0$). It is important to note that these characteristics for the $n$=0 LL are kept for many other values of disorder and magnetic fluxes investigated in this work.

The localization properties of the states within the $n$=0 LL are inspected, as shown in Fig.1(c), throughout the calculation of the participation ratio (PR)\cite{thouless}, defined by:

\begin{equation}
PR = \frac{1}{N' \sum_{i=1}^{N'}|\psi_{i}|^{4}},
\end{equation}

\hspace{-\parindent}where $\psi_{i}$ is the amplitude of the normalized wave function on site $i$, and $N'=M\times N$ is the total number of lattice sites. The PR gives therefore the proportion of the lattice sites over which the wave function is spread. In this way, the PR for a localized state vanishes in the thermodynamic limit, while peaks in the PR indicate the presence of extended states (critical energies).

Fig.1(c) allows the observation of two peaks in the PR, clearly indicating the splitting of two critical energies. Note that this splitting occurs exclusively for the $n$=0 LL in the considered model: the PR calculated for the higher levels (not shown here) shows always only one critical energy (one PR peak) per Landau band.
It is also important to point out that the splitting of the two critical energies is well defined even considering that there is no clear opening of an energy gap in the DOS (see section 6 for further discussion on this). As observed in the comparison between Fig.1(b) and Fig.1(c), there are states contributing to the DOS in the energy range between the two PR peaks, however, the closer these states are to E=0, the more localized they are, i.e., the smaller is their localization length.


\begin{figure}
\begin{center}
\includegraphics[width=10.3cm]{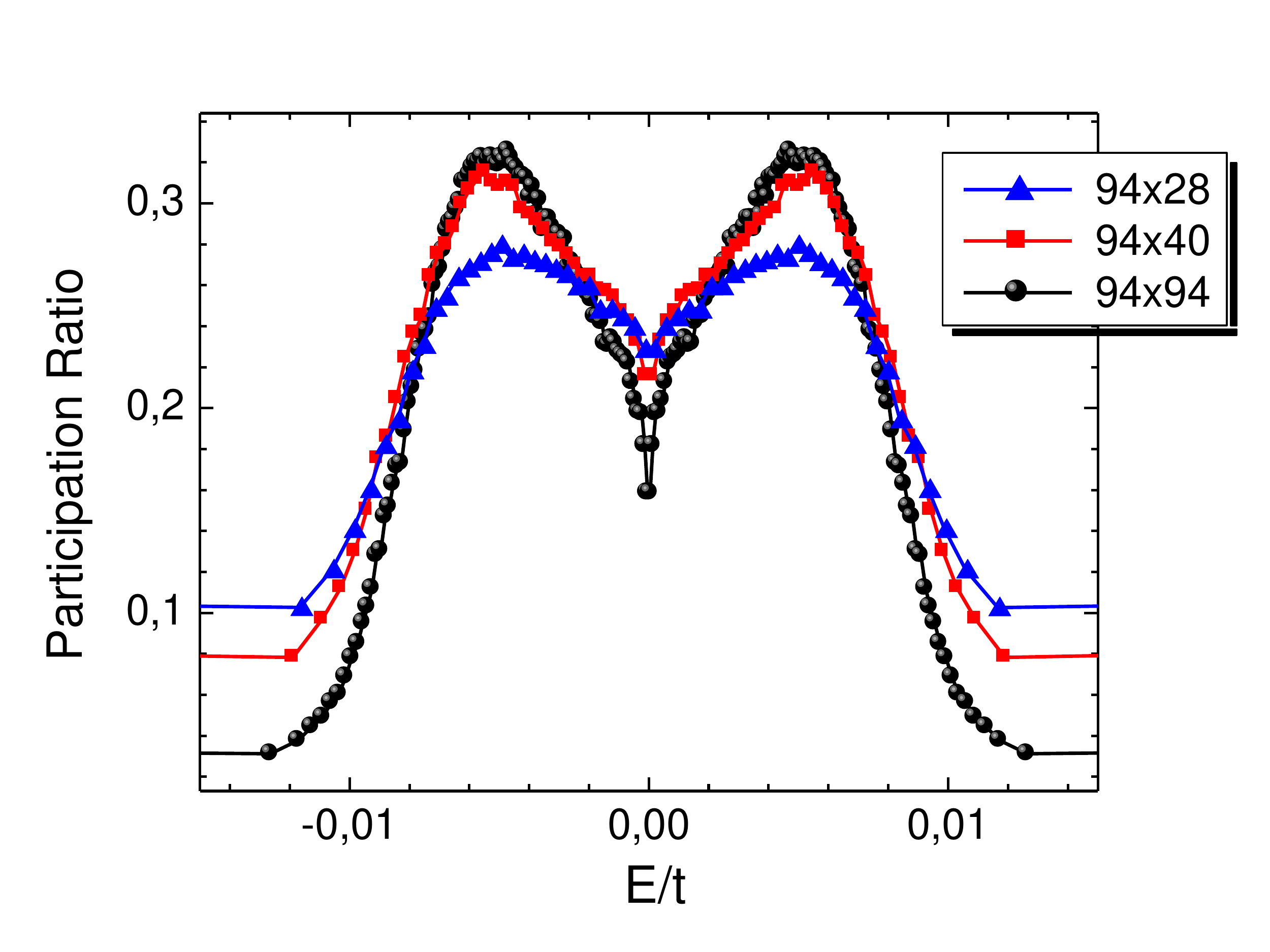}
\end{center}
\caption{PR for the states within the $n$=0 LL, for three different lattice sizes, showing how the peaks identifying the two critical energies get better definition as the system size is increased. The lattice sizes $M$$\times$$N$ ($L_x$$\times$$L_y$) and number of disorder realizations considered are: (i) 94$\times$94 (19.8nm$\times$11.4nm), 360 realizations;  (ii) 94$\times$40 (19.8nm$\times$4.8nm), 700 realizations; and (iii) 94$\times$28 (19.8nm$\times$3.3nm), 1000 realizations. For all the sizes, $W/t=0.1$ and $\Phi/\Phi_0=2/94\approx 0.021$.}
\end{figure}


Throughout this work, various PR calculations are shown, corresponding to different system sizes, different values of magnetic flux and different disorder strengths. In order to examine the question of possible finite lattice-size effects on the participation ratio calculations and get a better understanding of the PR peaks, in Fig.2 the PR is shown for three different lattice sizes. The dimension $L_x$=$19.8$nm is kept constant while $L_y$ varies from $L_y$=$3.3$nm, to $L_y$=$4.8$nm and to $L_y$=$11.4$nm  (more details are specified in the figure caption). Disorder and magnetic fluxes values are the same for the three cases: $W/t$=$0.1$ and $\Phi/\Phi_0$=$2/94$$\approx$$0.021$. The magnetic length for this flux is $l_B$= $\sqrt{\hbar/(eB)}$=$6.3$\AA. Fig.2 shows how the peaks in the PR are better defined as the system size is increased. As expected, the PR values corresponding to the localized states (at the LL tails and LL center) decrease with increasing system size \cite{thouless}.

The periodical boundary conditions impose restrictions over the values of magnetic flux in order to guarantee a commensurability between the wave function periodicity (the phases in the hopping) and the periodicity of the lattice considered. For a system with $M\times N$ sites, and considering the Landau gauge, the possible sequence of flux values is: $\Phi/\Phi_0=2/M,4/M,6/M,...$. Consequently, to consider small values of magnetic flux it is necessary to increase $M$ (i.e., increase the $L_x$ dimension), however, computational limitations restrict the maximum matrix sizes to be diagonalized. To analyze small fluxes it is then usually necessary to consider rectangular lattices, which are bigger in $L_x$ than in $L_y$. However, we have to keep in mind that $L_y$ should still be many times greater than $l_B$ in order to avoid finite size effects and to the PR peak to be well defined, as Fig.2 suggests. It is worth to note that the increase of $N$ in Fig.2 determines also the increase of the number of states per Landau level. The degeneracy is given by $d=M$$\times$$N$$\times$$\Phi/\Phi_0$ and, due to the flux restrictions discussed above, for a flux $\Phi/\Phi_0=2/M$, we have $d=2N$ states per LL.

\section{Probability Densities in each of the Sublattices}

\begin{figure}[b]
\vspace{0.7cm}
\includegraphics[width=15cm]{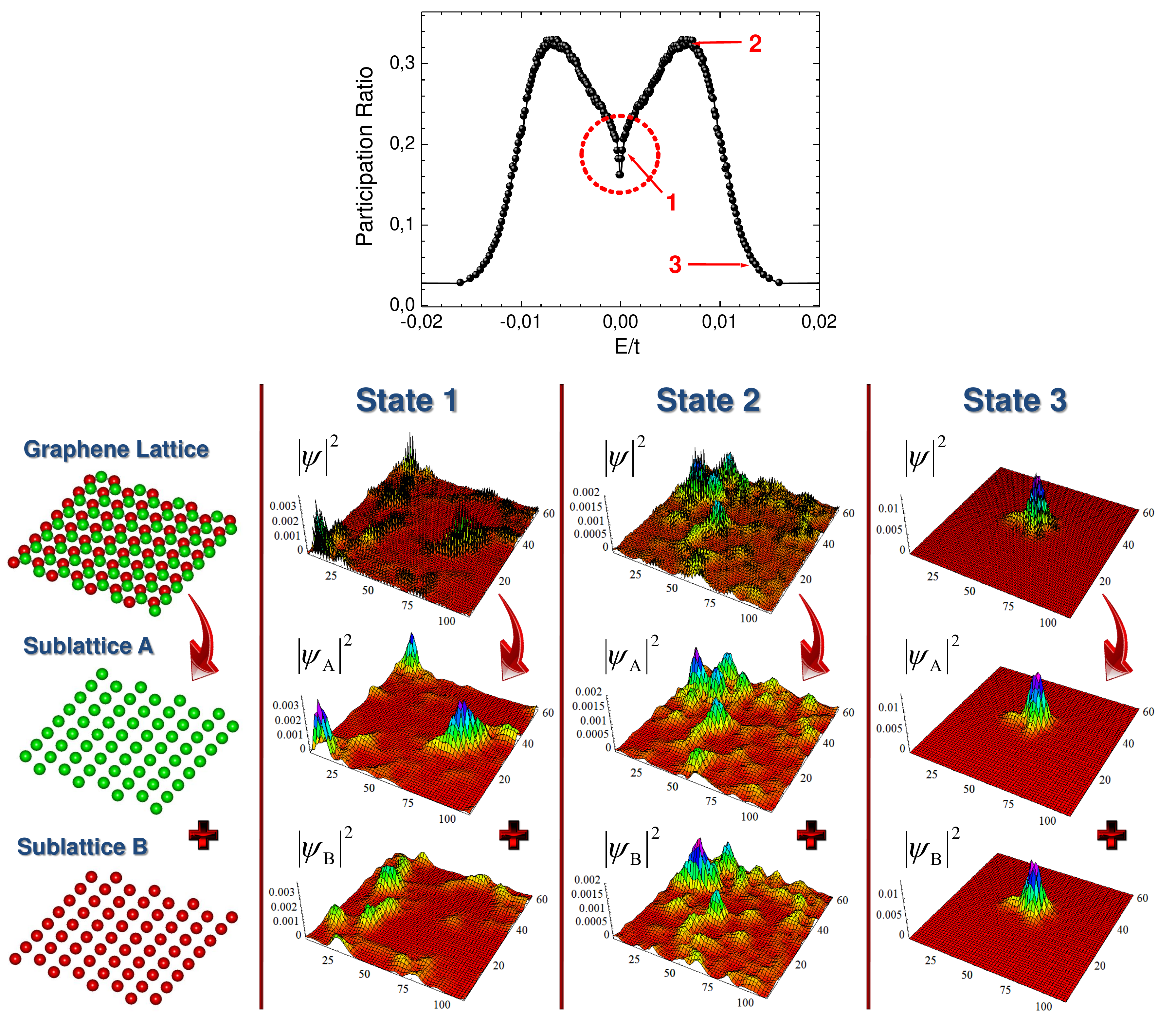}
\caption{{\bf (top)} Participation Ratio of the $n$=0 LL states from a graphene lattice with size $L_x$$\times$$L_y$=12.6nm$\times$13.2nm ($M$$\times$$N$=60$\times$108 atoms),  $\Phi / \Phi_0$=2/60$\approx$0.033, $W/t$=0.1 averaged over 500 disorder realizations. {\bf (bottom)} Probability densities ($|\Psi|^2$) and their decompositions, showing separately the contributions from each
sublattice ($|\Psi_A|^2$ and $|\Psi_B|^2$), for three selected states indicated by the arrows. While for the states 2 and 3 the probability density has the very same spatial distribution in each of the two sublattices, the state 1 is peculiar  because it has completely different and independent probability densities in each sublattice (all the states close to E=0, from the region inside the circle, show this behavior).}
\end{figure}

In order to try to get more information about the states within the $n$=0 LL and, consequently, about the origin of the critical energies splitting observed, the wave functions of these states are now examined. Special attention is paid to how the wave function amplitudes are distributed in each of the two sublattices, once in absence of disorder all the eigenstates within the $n$=0 LL have non-zero amplitudes only on one of the sublattices \cite{ando}.

In Fig. 3 it is shown the participation ratio calculated for a graphene lattice with $M$$\times$$N= 60$$\times$108 atoms ($L_x$$\times$$L_y=12.6$nm$\times$13.2nm), considering a magnetic flux  $\Phi / \Phi_0=2/60\approx0.033$ and a random hopping amplitude $W/t$=0.1 averaged over 500 disorder realizations. It is also shown, for three selected states indicated by the arrows, the wave function probability densities, $|\Psi|^2$, plotted over all the 60$\times$108 lattice sites, and then, bellow, the probability densities over each sublattice are plotted separately.  After calculating the eigenvectors from the exact Hamiltonian diagonalization, we have the probability density  $|\Psi_{(i,j)}|^2$ for all the lattice sites $(i,j)$. Then, the decomposition shown consists simply of plotting separately the probability densities for only sites $(i,j)$ which belong to one of the sublattices: $A$ ($|\Psi_A|^2$) or $B$ ($|\Psi_B|^2$). Each one of the three selected states corresponds to a different characteristic region from the $n$=0 LL: state 1 is a localized state very close to $E$=0, state 2 is a state from the region of critical (extended) states at the PR peak, and state 3 is a localized state from the band tail. 

The most interesting observation refers to the state 1, for which the spatial distribution of the probability density on one sublattice is completely different from that on the other one. Observing $|\Psi_A|^2$ and $|\Psi_B|^2$ separately, after the sublattice decomposition, one can see that for the state 1 the distribution of maxima and minima amplitudes on one sublattice occurs in different positions and is completely independent from the distribution on the other. All the states in the region delimited by the dashed circle,  states close to $E$=0, have this absolutely uneven spatial distribution of amplitudes between the sublattices. Observe that this is the region where the two Landau bands (split after the symmetry breaking) have a more significant overlap.
For the states in the region right after the limits of the circle, some position correlation between the two sublattices starts to appear in the probability density (not shown here). But when reaching the proximities of the PR peak, and after that until the band tail, we observe that all the states show a symmetric distribution between the sublattices, as can be observed from the decompositions of states 2 and 3: for these states, the amplitudes on one sublattice looks pretty much like the amplitudes on the other.

It is worth to note that although having  distributions that are asymmetric between the sublattices, it is not the case that the wave function is more concentrated in one sublattice than in the other. The behavior observed here is different from the observed when the disorder is introduced through on-site energy fluctuations (diagonal disorder models), for which the amplitudes over one sublattice are more significant than over the other \cite{ana_valley,ana_hmf18}.
In fact, both sublattices have equally significant amplitudes (note that the vertical scales of the graphics are kept the same in the decomposition): the sum of all amplitudes over the sublattice $A$ was calculated and is pratically equal (with differences smaller than 5$\%$) to the sum over the sublattice $B$.

Observing the wavefunction amplitudes in Fig. 3 one sees that the localization length of state 1 is larger than the one of state 3. Nevertheless, this difference reflects exactly the difference in the PR values of these states: for this system size, the PR for state 1 is not as small as for state 3 (remember however that the lattice size analysis shown in Fig. 2 indicated that both state 1 and state 3 are truly localized states). The reason for the larger localization length for the state 1 compared to state 3 is most probably due to the mixing of states in this region between the PR peaks, where the split LLs are overlapped (see section 6).

Another point to be noted is  that, for the disorder model considered (only off-diagonal, with random nearest-neighbor hoppings), there is a perfect symmetry around $E$=0: the probability density of the state at an specific energy $E$ is exactly equal to that of the state at $-E$. One can also observe this through  the perfect symmetry of the participation ratio curve around $E=0$.

\section{Splitting as a function of Disorder and Magnetic Field}

The evolution of the critical energies splitting in the $n$=0 LL with increasing disorder strength and magnetic field is investigated here, throughout the results shown in Figures 4 and 5, respectively.

\begin{figure}[b]
\vspace{0.5cm}
\includegraphics[width=15.3cm]{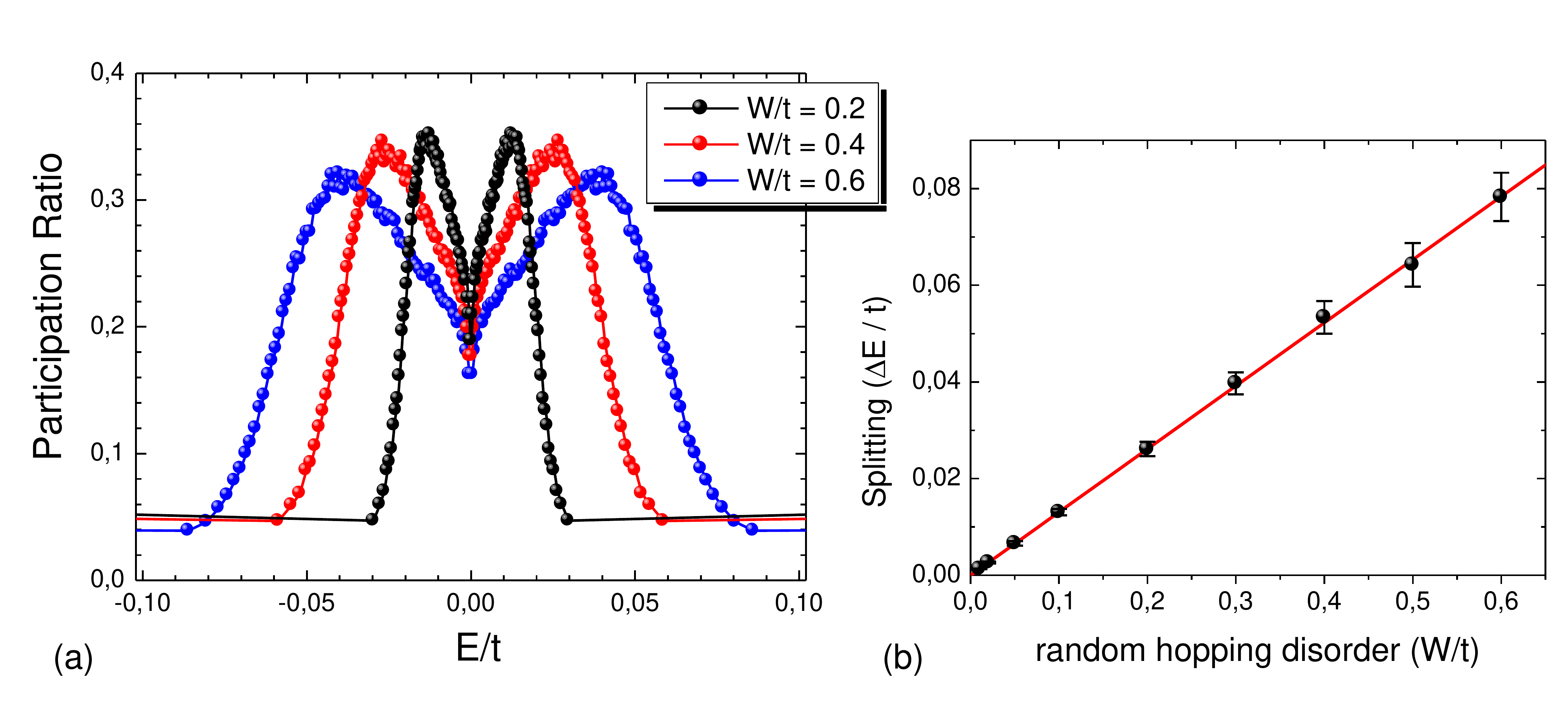} \hspace{-0.5cm}
\caption{{\bf (a)} Participation ratio for the states from the $n$=0 LL, for three different disorder strengths. The energy splitting between the two peaks is observed to increase with disorder. Lattice size is 13.9nm$\times$9.2nm ($M$$\times$$N$=66$\times$76) and the magnetic flux $\Phi / \Phi_0$=2/66$\approx$0.030.    {\bf (b)} Energy splitting of the two critical energies versus $W/t$, the random hopping disorder strength, showing linear fit.}
\end{figure}


Fig.4(a) shows the PR of the states from the $n$=0 LL, for three different disorder strengths: $W/t$=0.2, $W/t$=0.4 and $W/t$=0.6. One can see that the energy splitting $\Delta$$E/t$ between the two critical energies (between the two PR peaks) is increased as the disorder increases the Landau level broadening. In Fig.4(b) the splitting $\Delta$$E/t$ is plotted as a function of disorder, for several values of $W/t$. The linear regression fits well the data for the wide interval of disorder considered: the red line passes through origin and has angular coefficient 0.1306 $\pm$0.0006, which is valid for the specific magnetic flux considered here ( $\Phi / \Phi_0$=2/66$\approx$0.030): 

\begin{equation}
\frac{\Delta E}{t} = (0.1306 \pm 0.0006) \frac{W}{t},
\end{equation}

The results shown in Fig.4 are calculated for lattices having size 13.9nm$\times$9.2nm ($M$$\times$$N$=66$\times$76) and for a magnetic flux $\Phi / \Phi_0$=2/66$\approx$0.030, averaging the results for each $W/t$ over 400 disorder realizations.


\begin{figure}
\vspace{0.1cm}
\includegraphics[width=15.3cm]{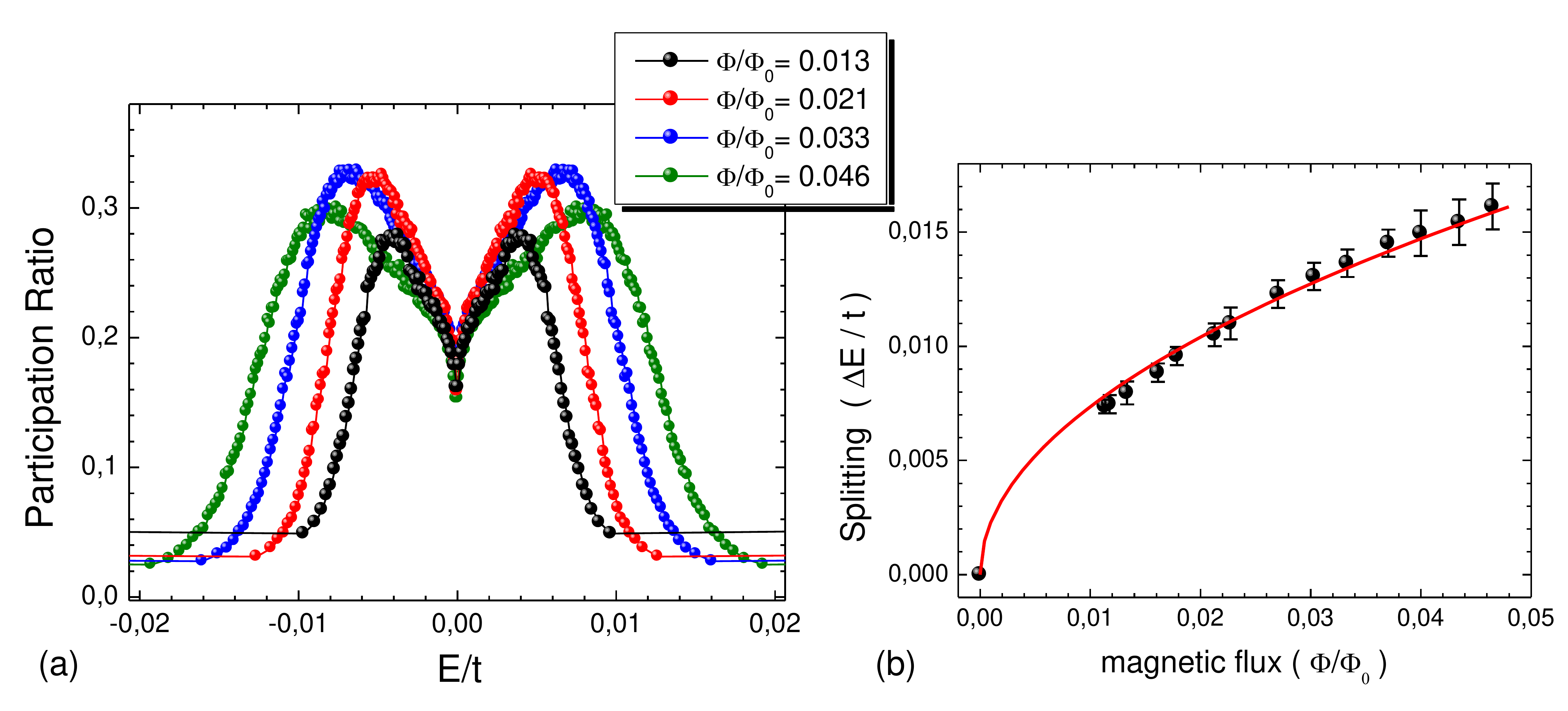}
\caption{{\bf (a)} Participation ratio for the states from the $n$=0 LL, for four different magnetic fluxes. The energy splitting between the two peaks is observed to increase with magnetic flux. {\bf (b)} Energy splitting of the two critical energies versus magnetic flux $\Phi / \Phi_0$, showing a square root dependence. The lattice dimensions vary for each flux considered and are listed in table 1.  The disorder strength is $W/t$=0.1 for all of the fluxes. }
\end{figure}


In Fig.5 the dependence of the splitting with magnetic field is examined. Fig.5(a) shows four examples of PR calculations within  the $n$=0 LL, for fixed disorder ($W/t=0.1$) and different magnetic fluxes. An increasing energy splitting of the two PR peaks is observed with increasing magnetic flux. In Fig.5(b) this splitting is plotted as a function os magnetic flux for several magnetic fluxes, and we observe that the data is well adjusted by a square root fit (the specific coefficient value is valid for the specific disorder $W/t$=0.1 considered):

\begin{equation}
\frac{\Delta E}{t} = (0.0736 \pm 0.0005) \sqrt{\frac{\Phi}{\Phi_0}}.
\end{equation}
 
The dimensions of the lattices considered for each flux, as well as the magnetic lengths and number of disorder realizations undertaken are listed in table 1.

\begin{table}
\caption{\label{label}Lattice dimensions, magnetic length $l_B$ and number of disorder realizations, for each of the magnetic fluxes considered to define the data points in Fig.5.}
\lineup
\begin{indented}
\item[]\begin{tabular}{@{}llllll}
\br
$\Phi / \Phi_0$ & $M$$\times$$N$ & lattice sites & $L_x$(\AA)$\times$$L_y$(\AA) & $l_B$(\AA) & realizations \\
\mr
0.0114 & 176$\times$50 & \08.800 & 372.7$\times$60.3 & 8.57 & 350\\
0.0118 & 170$\times$60 & 10.200 & 360.0$\times$72.6 & 8.42 & 210\\
0.0133 & 150$\times$62 & \09.300 & 317.4$\times$75.0 & 7.91 & 410\\
0.0161 & 124$\times$40 & \04.960 & 262.0$\times$48.0 & 7.19 & 390\\
0.0179 & 112$\times$44 & \04.928 & 236.4$\times$52.9 & 6.83 & 360\\
0.0213 & \094$\times$94 & \08.836 & 198.1$\times$114.4 & 6.26 & 360\\
0.0227 & \088$\times$56 & \04.928 & 185.3$\times$67.6 & 6.06 & 280\\
0.0270 & \074$\times$58 & \04.292 & 155.5$\times$70.1 & 5.55 & 400\\
0.0303 & \066$\times$76 & \05.016 & 138.5$\times$92.2 & 5.25 & 400\\
0.0333 & \060$\times$108 & \06.480 & 125.7$\times$131.6 & 5.00 & 500\\
0.0370 & \054$\times$110 & \05.940 & 112.9$\times$134.1 & 4.74 & 600\\
0.0400 & \050$\times$90 & \04.500 & 104.4$\times$109.5 & 4.56 & 280\\
0.0435 & \046$\times$130 & \05.980 & \095.8$\times$158.7 & 4.38 & 500\\
0.0465 & \086$\times$60 & \05.160 & 181.1$\times$72.6 & 4.23 & 400\\
\br
\end{tabular}
\end{indented}
\end{table}

Regarding the square root dependence of the splitting with magnetic field, observe that the critical energies from the $n$=0 LL follow then the same dependence with $B$ presented by the higher Landau levels.

It is important to note that these results are in agreement with reference \cite{schweitzer}, where a linear dependence of the energy splitting with the disorder, and a  square root dependence with the magnetic field were observed for a random magnetic-field disorder model, and where the splitting was determined by means of the two-terminal conductance peaks.

\vspace{1cm}

\section{Interplay between Splitting and LL Broadening}

It is observed in Fig.1 that the splitting is much better defined trough the localization properties of the states (definition of two clearly split PR peaks) than trough the density of states. Nevertheless, the DOS shape observed in Fig.1(b) suggests a superposition of two split bands. Indeed, using a multi-peak fitting procedure, shown in Fig.6 (a) and (b), it is found that the DOS obtained are reasonably well fitted by two overlapping Gaussian curves (apart from the small peak at $E=0$, whose origin is not understood). In Fig.6(c) it is possible to compare overall broadenings of the two $n$=0 LL DOS, which are calculated for two different values of disorder, $W/t$=0.05 and $W/t$=0.6, keeping constant all the other parameters: lattice size is $M$$\times$$N$=66$\times$76 and the magnetic flux considered is $\Phi / \Phi_0$=$2/66$=0.0303. Fig.6 (d) shows that two Gaussian peaks fitted to the DOS have a good coincidence with the calculated PR peaks positions.


\begin{figure}
\vspace{0.1cm}
\includegraphics[width=15.5cm]{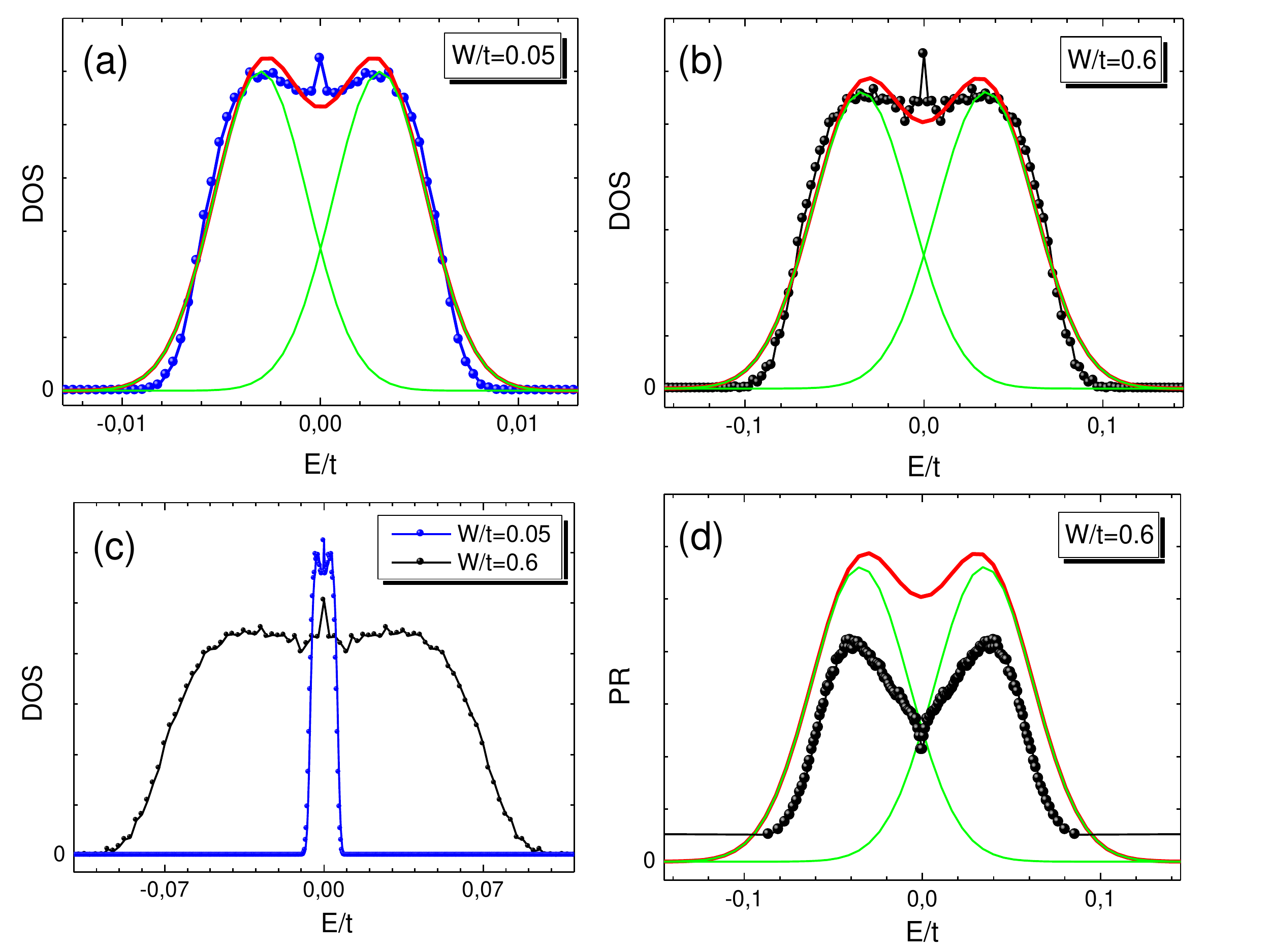}
\caption{Gaussian multi-peaks fitting of the DOS. {\bf (a)} The circles indicate the DOS calculated for the lattice with $M$$\times$$N$=66$\times$76 atoms, $W/t$=0.05 and $\Phi / \Phi_0$=0.0303. Green lines show the two Gaussian peaks fitted to this DOS, while the red line is the overall fitting (sum of the two peaks). {\bf (b)} Same as (a), now for $W/t$=0.6. {\bf (c)} DOS shown at the same scale for the $W/t$=0.05 and $W/t$=0.6. {\bf (d)} Participation Ratio (circles) and the fitting of the corresponding DOS for $W/t$=0.6.}
\end{figure}


The splitting is observed in the previous section, through the separations of the two PR peaks, to have a square root dependence with magnetic field, and to increases linearly with the disorder. However, what is important to discuss now is that, even when the splitting increases, the overlapping between the two degeneracy broken levels shows to be pretty much  about the same, for a wide range of parameters observed, as the comparison between Fig.6(a) and Fig.6(b) indicates. This is because the same increasing disorder or magnetic field that causes the increasing of the critical energies splitting, also causes the increasing broadening of each Landau band. In other words, we would need that the rate on which the splitting increases with disorder and with magnetic field were much higher than the rate on which the LLs get broadened by them, to be able to observe an increasingly well defined separation of the DOS in two independent (not overlapped) bands. The interplay between both of these effects (splitting and broadening of the bands) produces in the end an overall DOS for the $n$=0 LL having always the same appearance (the difference being only the broadening, but the main features are kept the same).  This indicates a direct connection between the effects of LL broadening and LL splitting.  To the experiments, one consequence of this behavior is that it might be equally difficult to observe a real energy gap throughout DOS measurements, independent of the amount of disorder or of the intensity of the magnetic field.

The random hoppings considered here are spatially uncorrelated. A further desirable extension for this work is to consider a spatial correlation for the hoppings, a situation that closer emulates the ripples in real graphene sheets. In Ref. \cite{aoki} it is observed that the introduction of an increasingly higher correlation length for the random hoppings in graphene results in an increasingly thinner $n$=0 LL compared to the broadening of the higher LLs. On the other hand, the splitting of two critical energies within the $n$=0 LL is observed in Ref. \cite{schweitzer}  for a random flux model, for which the $n$=0 LL is much thinner than the higher levels (even being thinner, the DOS shows a $n$=0 LL shape very similar to the observed here). The results of these two works \cite{schweitzer,aoki}, together with the results shown here, suggest that the splittings in the central LL should survive when a finite correlation length for the hoppings is taken into account in the model, however with the energy splittings being scaled down with the width of the Landau band, in a similar fashion to the observed in Fig. 6.

\section{Conclusions}

The breaking of the degeneracy of the graphene $n$=0 LL is investigated in this work through a simple numerical model, considering a non-interacting tight-binding model for a hexagonal lattice with random nearest-neighbors hoppings. Inferring the localization properties of the states by means of participation ratio calculations, two clearly split critical energies are observed. The origin for this splitting has to be intrinsic to the disorder model considered, which introduces inter-valley mixing to the system.

The energy splitting  is observed to have a linear dependence with the disorder strength, and a  square root dependence with the magnetic field, confirming the results obtained in ref. \cite{schweitzer}. The analysis of the probability densities of the wave functions within the $n$=0 LL shows that there is a region, for the states closer to $E$=0, where there is an important asymmetry in the distribution of the wave function amplitudes between the two sublattices. In this region, although there are amplitudes over both sublattices, there is no matching of the spatial positions of the amplitudes   over each sublattice. It is like as each sublattice has its own probability density, completely independent from the other. This may lead to a diminish in the state percolation over the lattice, increasing even further the localized character of these states and also influencing on anomalous localization properties. It is also observed that this region of states with special wave function distribution between the sublattices coincides with the region where the two degeneracy broken Landau levels have a more important overlapping. Another interesting observation is that this overlapping keeps approximately constant for different disorder strengths, due to the interplay between LL broadenings and LL splitting with increasing disorder or magnetic flux. These results may help in elucidating the physics involved in the splitting of the $n$=0 LL due to valley/sublattice symmetry breaking.


\vspace{0.5cm}

\section{Acknowledgements}

The author acknowledges fruitful discussions with Peter A. Schulz, Yakov Kopelevich, Caio H. Lewenkopf and Eduardo R. Mucciolo. This work is supported by FAPESP. Some of the numerical simulations were performed on the clusters from CENAPAD/UNICAMP and IFGW/UNICAMP.



\section{References}

\begin{thebibliography}{0}



\bibitem{novoselov} K. S. Novoselov, A. K. Geim, S. V. Morozov, D. Jiang, M. I. Katsnelson, I. V. Grigorieva, S. V. Dubonos, A. A. Firsov, Nature \textbf{438}, 197 (2005).

\bibitem{zhang1} Y. Zhang, Y.-W. Tan, H.L. Stormer, P. Kim, Nature \textbf{438}, 201 (2005).

\bibitem{reviews} A. K. Geim and K. S. Novoselov, Nat. Mat. \textbf{6}, 183 (2007); A. H. Castro Neto, F. Guinea, N. M. R. Peres, K. S. Novoselov and A. K. Geim , Rev. Mod. Phys. 81, 109 (2009).




\bibitem{zhang2} Y. Zhang, Z. Jiang, J. P. Small, M. S. Purewal, Y.-W. Tan, M. Fazlollahi, J. D. Chudow, J. A. Jaszczak, H. L. Stormer, and P. Kim, Phys. Rev. Lett. {\bf 96}, 136806 (2006).

\bibitem{jiang} Z. Jiang, Y. Zhang,  H. L. Stormer, and P. Kim, Phys. Rev. Lett. {\bf 99}, 106802 (2007).

\bibitem{ong} J. G. Checkelsky, L. Li and N. P. Ong, Phys. Rev. Lett. {\bf 100}, 206801 (2008).

\bibitem{andrei} G. Li, A. Luican and E. Y. Andrei, Phys. Rev. Lett. {\bf 102}, 176804 (2009).

\bibitem{giesbers} A. J. M. Giesbers,  L. A. Ponomarenko, K. S. Novoselov, A. K. Geim,  M. I. Katsnelson, J. C. Maan  and U. Zeitler, arXiv:0904.0948 (unpublished).



\bibitem{ando} M. Koshino and T. Ando, Phys. Rev. B {\bf 75}, 033412 (2007).

\bibitem{nomura1} K. Nomura and A. H. MacDonald, Phys. Rev. Lett. {\bf 96}, 256602 (2006).

\bibitem{lederer} J. N. Fuchs and P. Lederer, Phys. Rev. Lett. {\bf 98}, 016803 (2007).

\bibitem{lederer2} J. N. Fuchs and P. Lederer, Eur. Phys. J. Special Topics {\bf 148}, 151 (2007).

\bibitem{alicea} J. Alicea and M.P.A. Fisher, Solid State Commun. {\bf 143}, 504 (2007).

\bibitem{levitov} D. A. Abanin, P. A. Lee and L. S. Levitov, Phys. Rev. Lett. {\bf 98}, 156801 (2007).

\bibitem{yang} K. Yang, Solid State Commun. {\bf 143}, 27 (2007).

\bibitem{mirlin} P.M. Ostrovsky, I.V. Gornyi and A.D. Mirlin, Phys. Rev. B {\bf 77}, 195430 (2008).

\bibitem{schweitzer} L. Schweitzer and P. Markos, Phys. Rev. B {\bf 78}, 205419 (2008).

\bibitem{igor} I. A. Luk'yanchuk and A. M. Bratkovsky, Phys. Rev. Lett. {\bf 100}, 176404 (2008).

\bibitem{fertig} R. Cote, J. F. Jobidon and. A. Fertig, Phys. Rev. B {\bf 78}, 085309 (2008).

\bibitem{aoki} T. Kawarabayashi, Y. Hatsugai and H. Aoki, arXiv:0904.1927 (unpublished).

\bibitem{nomura2} K. Nomura, S. Ryu and D-H Lee, arXiv:0906.0159 (unpublished).

\bibitem{sarma} S. Das Sarma and K. Yang, arXiv:0906.2209 (unpublished).

\bibitem{ana_valley} A. L. C. Pereira and P. A. Schulz, Phys. Rev. B {\bf 77}, 075416 (2008).

\bibitem{ana_hmf18} A. L. C. Pereira and P. A. Schulz, Int. J.  Mod. Phys. B {\bf 23}, 2618 (2009).


\bibitem{thouless} D. J. Thouless, Phys. Reports {\bf 13}, 93 (1974).

\end{thebibliography}
\end{document}